\documentclass[final,5p,times,twocolumn]{elsarticle}
\usepackage{amssymb}
\usepackage{amsthm}
\usepackage{graphicx}
\usepackage{epsfig}
\journal{Gravitation and Cosmology}
\begin{document}

\begin{frontmatter}

\title{Cosmic acceleration a new review}

\author{O.\,A.\,Lemets}\ead{oleg.lemets@gmail.com}
\author{D.\,A.\,Yerokhin}\ead{denyerokhin@gmail.com}

\address{Akhiezer Institute for Theoretical Physics,
National Science Center "Kharkov Institute of Physics and
Technology", Akademicheskaya Str. 1, 61108 Kharkov, Ukraine}
\date{\today}

\begin{abstract}
Recent observations of near supernova show that the acceleration expansion of Universe decreases. This phenomenon is called the transient acceleration. In the second part of work we consider the 3-component Universe composed of a scalar field, interacting with the dark matter on the agegraphic dark energy background. We show that the transient acceleration appears in frame of such a model. The obtained results agree with  the latest  cosmological observations, namely, the 557 SNIa sample (Union2) was released by the Supernova Cosmology Project (SCP) Collaboration.
\end{abstract}

\end{frontmatter}

\section{Introduction}
At the begining of 21  century, the standard cosmological model (SCM) has become the dominant model of the universe, replacing the hot model of the universe (Big Bang). SCM is based on two important observational results: accelerated expansion of the universe and the Euclidean spatial geometry.
 In addition, it is assumed that the early universe is adequately described by the theory of inflation. SCM fixes a number of parameters of the universe and, in particular, its energy structure. According to the SCM  the Universe is currently dominated by dark energy (in the form of a cosmological constant $\Lambda$), required to explain the accelerated expansion and dark (non-baryonic) matter (DM). Existence of DM gives a possibility to solve a number of contradictions in the Big Bang model (non-decreasing behavior of rotation curves, the structure of galactic halo, a chronology of the  structures formation, etc.).

Attributing the acceleration of the universe expansion exclusively to the negative pressure generated by the cosmological constant drastically reduced the possibilities  of the scale factor dynamics and condemned  the universe to eternal accelerated expansion.

 In fact, the dynamics of the scale factor in SCM is described by Friedmann equations
\begin{equation}
\left( {\frac{{\dot a}}
{a}} \right)^2  = H_0^2 \left[ {\Omega _{m0} \left( {\frac{{a_0 }}
{a}} \right)^3  + \Omega _\Lambda  } \right]
\end{equation}
The solution of this equation reads

\begin{eqnarray}
  a(t) = a_0 \left( {\frac{{\Omega _{m0} }}
{{\Omega _{\Lambda 0} }}} \right)^{1/3} \left[ {sh\left( {\frac{3}
{2}\sqrt {\Omega _{\Lambda 0} } H_0 t} \right)} \right]^{2/3} ,  \\
  a\left( {t \ll H_0^{ - 1} } \right) \propto t^{2/3} ;\;a\left( {t \gg H_0^{-1}} \right) \propto e^{H_0 t} . 
\end{eqnarray}
We see that the asymptotic behavior of the solution described the era of matter domination in the early Universe $t\ll H_0^{-1}$ and dark energy domination for the later evolution $t\gg H_0^{-1}$. 

We now find the dependence of the deceleration parameter $q\equiv -a\ddot{a}/\dot{a}^2 = -\ddot{a}/\left( aH^2\right)$ on the redshift $z$ for a universe filled with arbitrary components of the state equation $p_i = w_i\rho_i$. In this case,
\begin{equation}
q = \frac{3}
{2}\frac{{\sum\limits_i  \Omega _i^{(0)} \left( {1 + w_i } \right)(1 + z)^{3\left( {1 + w_i } \right)} }}
{{\sum\limits_i  \Omega _i^{(0)} (1 + z)^{3\left( {1 + w_i } \right)} }} - 1
\end{equation}
For the SCM, this expression takes the form (see Figure 1)
\begin{equation}
q = \frac{1}
{2}\frac{{\Omega _M^{(0)} (1 + z)^3  - 2\Omega _\Lambda ^{(0)} }}
{{\Omega _M^{(0)} (1 + z)^3  + \Omega _\Lambda ^{(0)} }}
\end{equation}
In particular, at the present time

\begin{equation}
q_0  = \frac{{1 - 3\Omega _{\Lambda 0} }}{2}\simeq  - 0.6.
\end{equation}

A characteristic feature of the dependence $q(z)$ - a monotonous tendency to a limiting value of $q(z) = -1$ for $z \to -1$. Physically, this means that once the dark energy became dominant component (at $z \sim 1$), the universe in SCM is doomed to eternal accelerated expansion. Questioned the adequacy of this result is expressed repeatedly \cite{Andreas}, \cite{Barrow}.

\section{Transient acceleration: heuristic arguments}

Discovery of the transient acceleration is quite logical step to ensure the achievement of observational cosmology.  The first step on this thorny path can be regarded as a model of stationary universe. To accomplish the second step, and discover the Hubble expansion of the universe it took almost two thousand years.
The next step, though took far less time, still met a lot of objection: even now not every body agree with what was discovered to the accelerated expansion of the universe. We suggest  that cosmology now came close to make the next step in this direction. 

In this section, we discuss some considerations, both theoretical and observational, indicating that this path.

\subsection{Theoretical background}
J. Barrow \cite{Barrow} was one of the first to rise the question of validity of such  scenario. He showed that  in many well-motivated scenarios the observed period of vacuum domination is only a transient phenomenon. Soon after acceleration starts, the vacuum energy anti-gravitational properties are reversed, and a matter-dominated decelerating cosmic expansion resumes. Thus, contrary to general expectations, once an acceleration universe does not mean  an accelerating forever

To show this, we followed \cite{Barrow} considering a homogeneous and isotropic Universe with zero spatial curvature that contains two dominant forms of matter: a perfect fluid with pressure $p$ and density $\rho $ linked by an equation of state $p=w\rho $, with $w$ constant, together with a scalar quintessence field $\varphi $ defined by its self-interaction potential $V(\varphi)$.

Many theorists believe that fields with potentials of the form
\begin{equation}
V(\varphi) = V_p(\varphi) e^{-\lambda \varphi}.
\label{expVp}
\end{equation}
are predicted in the low energy limit of $M$-theory , where $V_p(\varphi)$ is a polynomial.
Albrecht and Skordis \cite{Andreas} have proposed a particularly attractive model of quintessence. It is driven by a potential which introduces a small minimum to the exponential potential, which is provided by the simplest  polynomial
\begin{equation}
V_p(\varphi) = (\varphi -\varphi_0)^{2} + A.
\label{Vp}
\end{equation}
and the potential takes the form
\begin{equation}
V(\varphi )=e^{-\lambda \varphi }\left( A+(\varphi -\varphi_0)^2\right),  \label{C}
\end{equation}  
where the constant parameters, $A$ and $\varphi_0$, are of order $1$ in Planck units, so there is also no fine tuning of the potential.
In this quintessence models, late-time acceleration is achieved
without fine tuning of the initial conditions.
Acceleration begins when the field gets trapped in the local minimum of the potential at 
$\phi =\phi _0+(1\pm \sqrt{1-\lambda ^2A})/\lambda $,
which is created by the quadratic factor in eq. (\ref{C}) when $1\geq\lambda ^2A$.
Once the field gets stuck in the false vacuum its kinetic energy disappears ($\phi \approx $ constant), and the ensuing dominance of $\rho +\rho _\phi $ by an almost constant value of the potential value triggers a period of accelerated expansion that never ends. 
In the article \cite{Barrow} it was found that this type of behaviour is by no means generic.

Transient vacuum domination arises in two ways. When $A\lambda ^2<1,$ the $\varphi $ field arrives at the local minimum with enough kinetic energy to roll
over the barrier and resume descending the exponential part of the potential
where $\varphi \gg\varphi _0$. This kinetic energy is determined by the scaling regime, and so by parameters of the potential and not by initial conditions.
Another instance of transient vacuum domination is the whole region $A\lambda ^2>1$. As $A$ increases towards $\lambda ^{-2}$, the potential
loses its local minimum, and flattens out into a point of inflexion. This is
sufficient to trigger accelerated expansion temporarily, but the field never
stops rolling down the potential, and matter-dominated scaling evolution
with $a(t)\propto t^{2/3}$ is soon resumed.
It is possible for the universe to exit from a period of accelerated expansion and resume decelerated expansion. Moreover, for the well-motivated family of Albrecht-Skordis potentials this is the most likely form of evolution, rather than a state of continuing acceleration.

\subsection{Observational evidence}
A. Starobinsky \cite{Starobinsky} and co-workers investigated the course of cosmic expansion in its  recent past using the Constitution SN Ia sample (which includes CfA data at low redshifts), jointly with
signatures of baryon acoustic oscillations (BAO) in the galaxy distribution
 and fluctuations in the cosmic microwave background (CMB). Allowing the equation of state of dark energy (DE) to vary, they find that a coasting model of
the universe ($q_0=0$) fits the data about as well as $\Lambda$CDM. This effect, which is most clearly seen using the recently
introduced $Om$ diagnostic \cite{Starobinsky_Om}, corresponds to an increase of
$Om$ and $q$ at redshifts $z\lesssim 0.3$. In geometrical terms, this suggests
that cosmic acceleration may have already peaked and that we are currently witnessing its slowing down. 

The case for evolving DE strengthens if a subsample of the Constitution set consisting of SNLS+ESSENCE+CfA SN Ia data is analysed in combination with BAO+CMB using the same statistical methods. 

\begin{figure}[t]
\centering
\psfig{figure=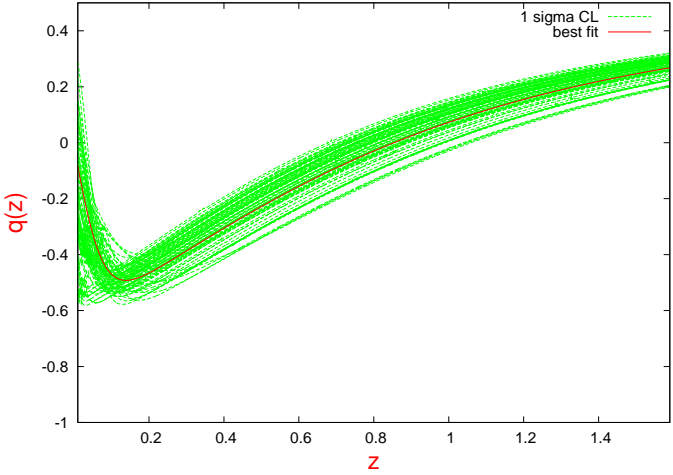,width=0.35\textwidth,angle=-90}
\caption{ The cosmological deceleration parameter $q(z)$  reconstructed using a combination of SN Ia, BAO and CMB data and the
ansatz (\ref{eq:step}).
Solid red lines show best fit reconstructed results
while dashed green lines show reconstructed results within $1\sigma$ CL\cite{Starobinsky} .}
\label{fig_Srarob}
\end{figure}

Note also that in article \cite{Starobinsky} it  was shown the impossibility to  come to agreement with   data obtained in  observations SN1a about CMB by  using the ansatz CPL. 
 To make this possible, they proposed a new ansatz
\begin{equation}
    w(z)=- \frac{1+ \tanh\left[(z-z_t)\Delta\right]}{2}.
\label{eq:step} 
\end{equation}
This fit ensures $w = -1$ at early times, and then increases the EOS
to a maximum of $w\sim 0$ at low $z$.
Figure\ref{fig_Srarob} shows the deceleration parameter $q$ and the $Om$ diagnostic reconstructed using (\ref{eq:step}).

 In the work \cite{LiWuYu,LiWuYu1}  for SNIa data,there was used the latest Union2 compilation released by the Supernova Cosmology Project (SCP) Collaboration recently \cite{Amanullah}. It consists of 557 data points and is the largest published SNIa sample up to now (2010 year).

 The authors  find that, independent of whether or not the systematic error
is considered, there exists a tension between low redshift data (SNIa+BAO) and high
redshift ones (CMB), but for the case with the systematic error considered this tension is weaker than  without the SNIa. By reconstructing the curves of q(z) and Om(z) from Union2+BAO, we obtain that for both the SNIa with and without the systematic error the cosmic acceleration has already peaked at redshift $z \sim 0.3$ and is decreasing. 
However, when the CMB data is added in  analysis, this result changes dramatically and the observation favors a cosmic expansion with an increasing acceleration,  indicating a tension between low redshift data and high redshift.

They  find that two different subsamples+BAO+CMB give completely different results on the cosmic expansion history when the systematic error is ignored, with one suggesting a decreasing cosmic acceleration, the other just the opposite, although both of them alone with BAO support that the cosmic acceleration is slowing down.

\begin{figure}[t]
\centering
\includegraphics[width=0.45\textwidth]{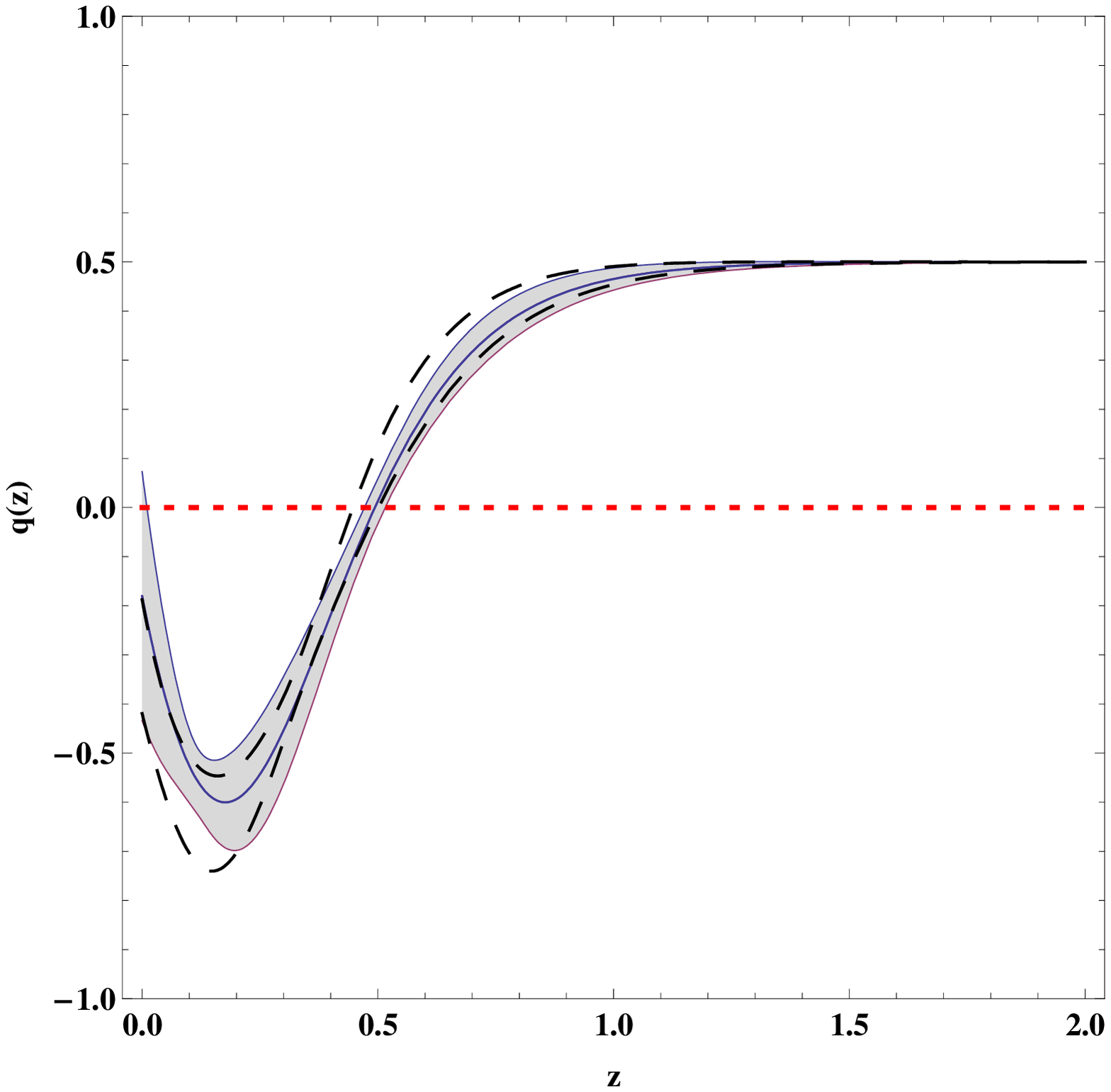}
\includegraphics[width=0.45\textwidth]{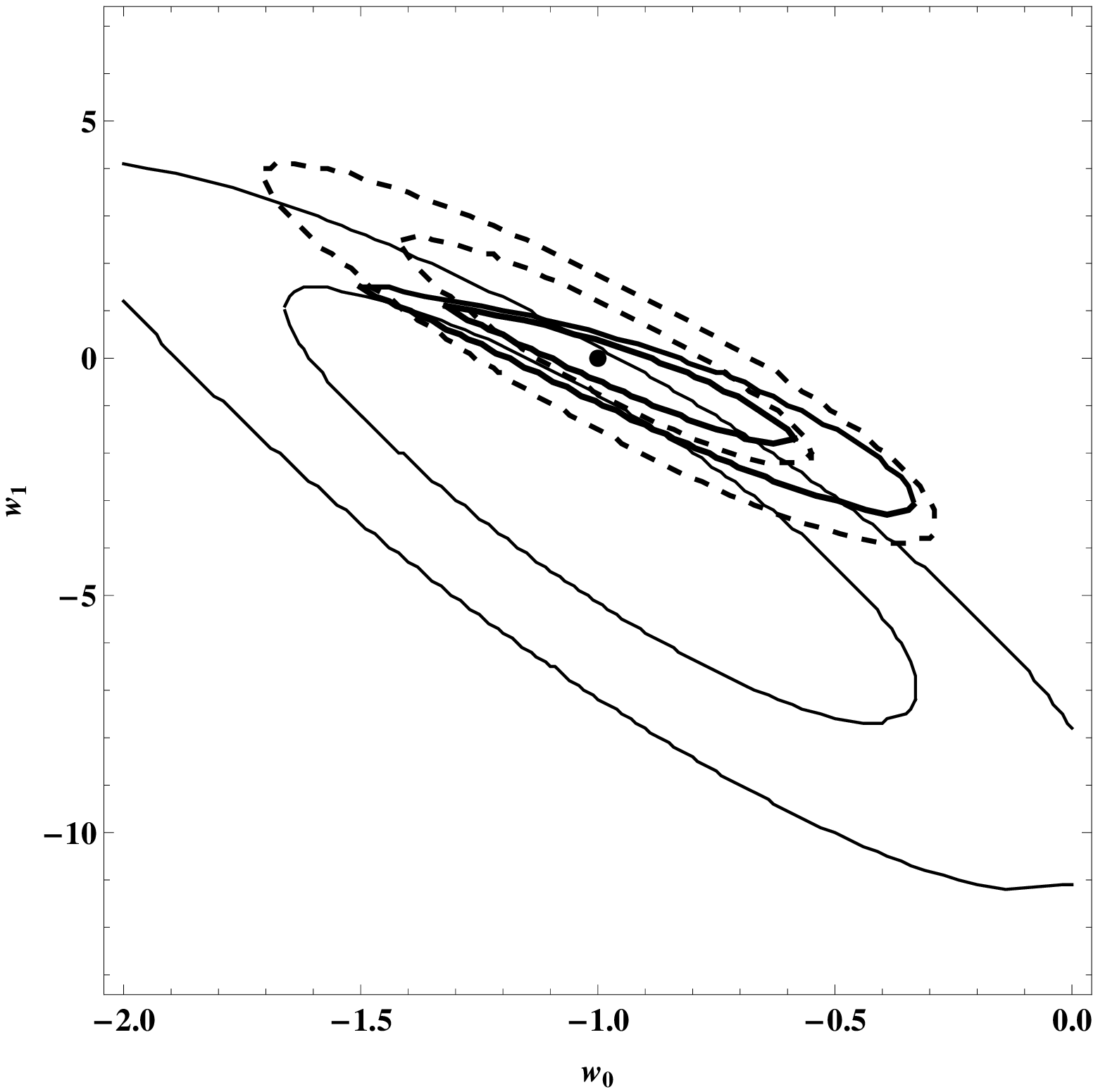}
\caption{At upper panel represent the results reconstructed  from Union2+BAO and show the evolutionary behaviors of $q(z)$  at the $68.3\%$
confidence level. The gray regions and the regions between two long
dashed lines  show the results without and with the systematic
errors in the SNIa, respectively. 
At lower panel represent the $68.3\%$ and $95\%$ confidence level regions
   for $w_0$ versus $w_1$ in the CPL parametrization,  $w=w_0+w_1z/(1+z)$.  In  the right panel, the system error in the SNIa is considered.  The dashed, solid   and
   thick solid lines represent the results obtained from Union2S, Union2S+BAO and Union2S+BAO+CMB, respectively. The point at $w_0=-1$, $w_1=0$ represents the spatially flat $\Lambda$CDM model. }
\label{fig:TrAcCh}
\end{figure}
Thus, the evolutional behavior of dark energy reconstructed and the issue of whether
the cosmic acceleration is slowing down or even speeding up is highly dependent upon the SNIa data sets, the light curve fitting method of the SNIa, and the parametrization forms of the equation of state. In order to have a definite answer, we must wait for data with more precision and search for the more reliable and efficient methods to analyze these
data.

\section{The model of interacting dark energy with a transient acceleration phase}\label{ISFA_TA}
In this section, we  investigate a new quintessence scenario driven by a rolling homogeneous scalar field with exponential potential $V(\varphi)$ interacting with dark matter on the agegraphic background. This  scenario   predicts  transient accelerating phase.

To describe the dynamic properties of such a Universe will adapt the system of equations (\ref{sys_xyz}) for this model. We introduce the modified variables:
\begin{eqnarray}  
x=\frac{\dot{\varphi}}{\sqrt{6}M_pH},\quad
y=\frac{1}{M_pH}\sqrt{\frac{V(\varphi)}{3}},\quad z=\frac{1}{M_pH}\sqrt{\frac{\rho_{m}}{3}},\quad  u=\frac{1}{M_pH}\sqrt{\frac{\rho_{_{q}}}{3}}.\label{var2}
\end{eqnarray}

The evolution of the scalar field is described by the Klein--Gordon equation, which in the presence of matter couplings is given by
\begin{equation}
\ddot{\varphi}+3 H \dot{\varphi} + \frac{dV}{d\varphi}=-  \frac{Q}{\dot{\varphi}}.
\label{eq:kgeqn}
\end{equation}
In this  section, we consider interactions $Q$ that are linear combinations of the scalar field and pressureless matter:
\begin{equation}
  \label{Q1}
  Q= 3H(\alpha\rho_\varphi+\beta \rho_m),
\end{equation}
where $\alpha$, $\beta$  are  constant parameters.

Without specifying the potential of the scalar field $V(\varphi),$ we obtain the system of equations in the form

\begin{eqnarray} 
x'&=& \frac{3x}{2}g(x,z,u)-3x + \sqrt{\frac{3}{2}}\lambda y^2-\gamma,\nonumber\hfill\\
 \label{sys_xyz}
y'&=&  \frac{3y}{2}g(x,z,u)- \sqrt{\frac{3}{2}}\lambda xy, \hfill\\
z'&=& \frac{3z}{2}g(x,z,u)-\frac{3}{2}z  + \gamma\frac{x}{z},\nonumber\hfill\\
u'&=& \frac{3u}{2}g(x,z,u)-\frac{u^2}{n},\nonumber\hfill
\end{eqnarray}

where 
$
g(x,z,u) = 2x^2+ z^2+ \frac{2}{3n}u^3
$
and
\begin{equation}
\gamma\equiv  -\frac{Q}{\sqrt{6}M_pH^2\dot{\varphi}},~\lambda\equiv  -\frac{1}{V}\frac{dV}{d\varphi} M_p.
\end{equation}
In these variables, we obtain

\begin{eqnarray}
  \nonumber
    Q&=&9H^3M_p^2\left[\alpha(x^2+y^2)+\beta z^3\right],\\ 
\gamma& = &\frac{\alpha(x^2+y^2)+\beta z^3}{x} \label{Q_gamma_xyz}. 
\end{eqnarray}

We will consider a scalar field with an exponential potential energy
density 
\begin{equation}\label{V_exp}
    V=V_0\exp\left(\sqrt{\frac{2}{3}}\frac{\mu\varphi}{M_p}\right),
\end{equation}
where $\mu$ is a constant.
In this case we obtain
%\begin{widetext}
\begin{eqnarray} 
x'&=& \frac{3x}{2}\left[g(x,z,u)- \frac{\alpha(x^2+y^2)+\beta z^2}{x^2}\right]-3x - \mu y^2,\nonumber\hfill\\
 \label{sys_xyz_V}
y'&=&  \frac{3y}{2}g(x,z,u)+\mu xy, \hfill\\
z'&=& \frac{3z}{2}\left[g(x,z,u)+\frac{\alpha(x^2+y^2)+\beta z^2}{z^2}\right]-\frac{3}{2}z ,\nonumber\hfill\\
u'&=& \frac{3u}{2}g(x,z,u)-\frac{u^2}{n}.\nonumber\hfill
\end{eqnarray}
%\end{widetext}
In this model the deceleration parameter has the form
\begin{equation}
q=-1+\frac 32\left[2x^2+z^2+\frac{2}{3n}u^3\right]. 
\end{equation}
Note that in this model, the cosmological parameters  are not  explicitly  depend of the parameters of interaction, but only determine by the behaviors of dynamical variables. This fact complicates the analysis of our model. 

The fact that $q$ is independent of the interaction term implies  that the region of phase space is the same for all of the models considered. 
It is possible to make some qualitative comments about the system
(\ref{sys_xyz_V}) for some special cases.

Initially, considering the case, $y = 0,$  we  obtain the constraining relation, which should be imposed on the parameters of interaction for the emergence of critical points. It is easy to show that in this case, the condition of real energy density requires
\begin{equation}
\label{cond}
    2\sqrt{\frac{\beta}{\alpha}}>1+\alpha+\beta,
\end{equation}
which necessarily requires $0>\beta>\alpha,|\alpha|+|\beta|<1.$ 
This critical point corresponds to the matter-dominated Universe and is unstable. The cases of several such points are possible, but they are of no interest.
Finally, it can be shown that any of this equilibrium point within (but not on) the boundary will exist for $x_0<0.$  For $z\neq 0$ and constrain satisfied (\ref{cond}), it
\begin{equation}
    x_c=\left[\left(a+\sqrt{{\beta}/{\alpha}}\right)^{1/2}+a\right]z_c,
\end{equation}
 where $a=\left(2\sqrt{{\beta}/{\alpha}}-(1+\alpha+\beta)\right)^{1/4}.$

\subsection{Review of the case $Q=0$}
In this subsection we  consider in more detail the case of  absence of interaction between the scalar field and dark matter.
Critical points of the system (\ref{sys_xyz_V}) in this $(\alpha=\beta=0)$ case are given in the Table \ref{tab:2}.
As can be seen from the table, the system (\ref{sys_xyz_V}), there are six physically admissible critical points, the latter of which is the attractor.
The first critical point, $(1,0,0,0),$ is unstable and  corresponds to the scalar field  dominated era with extremely rigid equation of state, the second critical point corresponds to the period of evolution when the scalar field behaves as a cosmological constant.
The other point, $(0,0,1,0),$ is physically unrealistic. It corresponds to a Universe filled with  dark matter (it contains neither scalar field, nor agegraphic dark energy) and it is also unstable.
Fourth critical point $(0,0,0,1)$  correspond to the Universe consisting  only of the  agegraphic dark energy and has already been discussed in detail. 

The physical interest present the final sixth critical point, which is the an attractor. It corresponds to the Universe consisting of a scalar field and agegraphic dark energy.
The location of this critical point is completely determined by the parameter of the potential $\mu$ and the value of $n:$
\begin{equation}
    \begin{array}{cccccl}
   x_{*}&=&\frac{2}{3n\mu}u_{*}, \quad& y_{*}&=&\sqrt{1-\left(1+\frac{4}{9n^2\mu^2}\right)u_{*}^2},\\  
	z_{*}&=&0, \quad						& u_{*}&=&\frac{3}{2n\mu^2}\left(-1+\sqrt{1+\frac{4n^2\mu^4}{9}}\right).
\end{array}
\end{equation}

The fact that $x_*\propto u_*$ is the characteristic feature of the tracking solutions. Note also that between the scalar field and agegraphic dark energy, evidence of background interactions. There is because the dynamics of the scalar field affects agegraphic dark energy. This effect is that agegraphic dark energy, having a negative pressure, affect on the rate of expansion of the Universe, which affects the Hubble parameter, which is included in the Klein-Gordon equation for a scalar field.

\begin{table}
\caption{Location of the critical points of the autonomous system of Eqs. (\ref{sys_xyz_V}), their stability and dynamical behavior of the Universe at those points.}
\label{tab:2}
\begin{tabular}{ccccc}
\hline\noalign{\smallskip}
$(x_c,y_c,z_c,u_c)$ & Stability & $q$ & $w_{\varphi}$& $w_{tot}$\\
 coordinates &  character  & & & \\
\noalign{\smallskip}\hline\noalign{\smallskip}
$\;(1,0,0,0)$ &  unstable  & $2$ &$ 1$& $1$\\
$\;(0,1,0,0)$   & unstable  & $-1$ &$-1$& $-1$\\
 $\;(0,0,1,0)$ &unstable &$\frac{1}{2}$\quad &$\nexists$&0\\
 $\;(0,0,0,1)$ & stable  & $-1+\frac{1}{n}$ &$ \nexists$& $-1+\frac{2}{3n}$\\
 $(-\frac{3}{2\mu},\frac{3}{2\mu},\sqrt{1-\frac{3}{2\mu^2}},0)$ & unstable &\quad$\frac{1}{2}$ &$\nexists$&0\\
 $(x_*,y_*,0,u_*)$ & attractor &$q_*<0$&$w_{\varphi*}$&$w_{tot*}$\\
\noalign{\smallskip}\hline
\end{tabular}
\end{table}

The attractor, once reached, brings to zero the matter density. To allow for the observed matter content of the Universe, we have to select the initial conditions, if they exist, in such a way that the attractor is not yet reached at the present time, but the expansion is already accelerated.

For example, consider the case when $\mu = -3, n=3.$ The dynamics of such a Universe is consistent with the observed data, which will be shown in the following section.  We note only that in this model the values of deceleration parameter and the scalar field state parameter in the attractor are equal respectively $q_*\approx -0.68,~w_{\varphi*}\approx -0.78.$

\subsection{Review of the case $Q=3H\alpha\rho_\varphi$}
In the above case, the phenomenon of transient acceleration that occurs in such a Universe does not match the observations.
However, it is possible to fit the observational data with this model in the case when dark matter and scalar field interact. In this subsection we consider the special case of interaction (\ref{Q1}) when $ \beta = 0.$

\begin{figure}[t]
\centering
\includegraphics[width=0.45\textwidth]{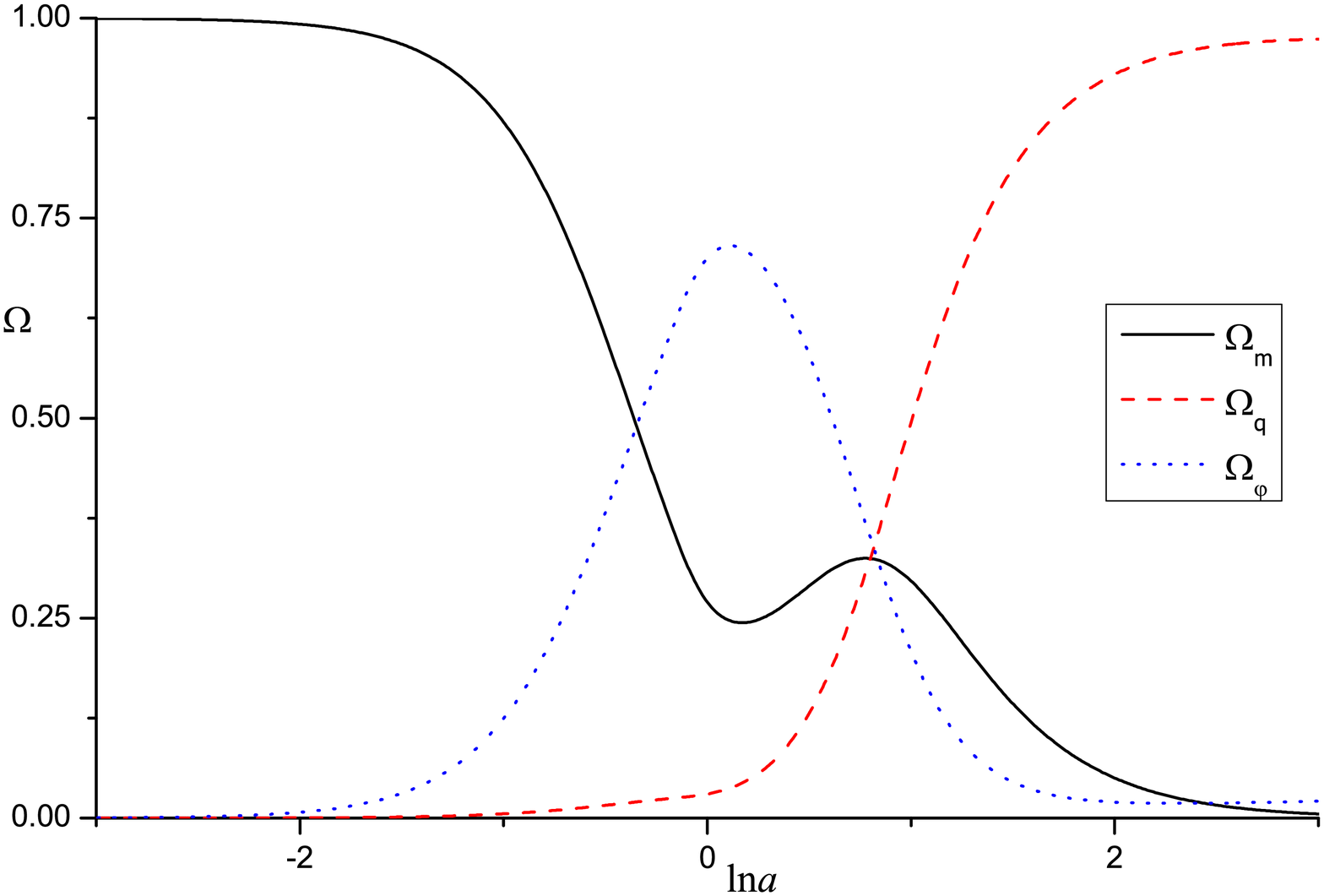}
\includegraphics[width=0.45\textwidth]{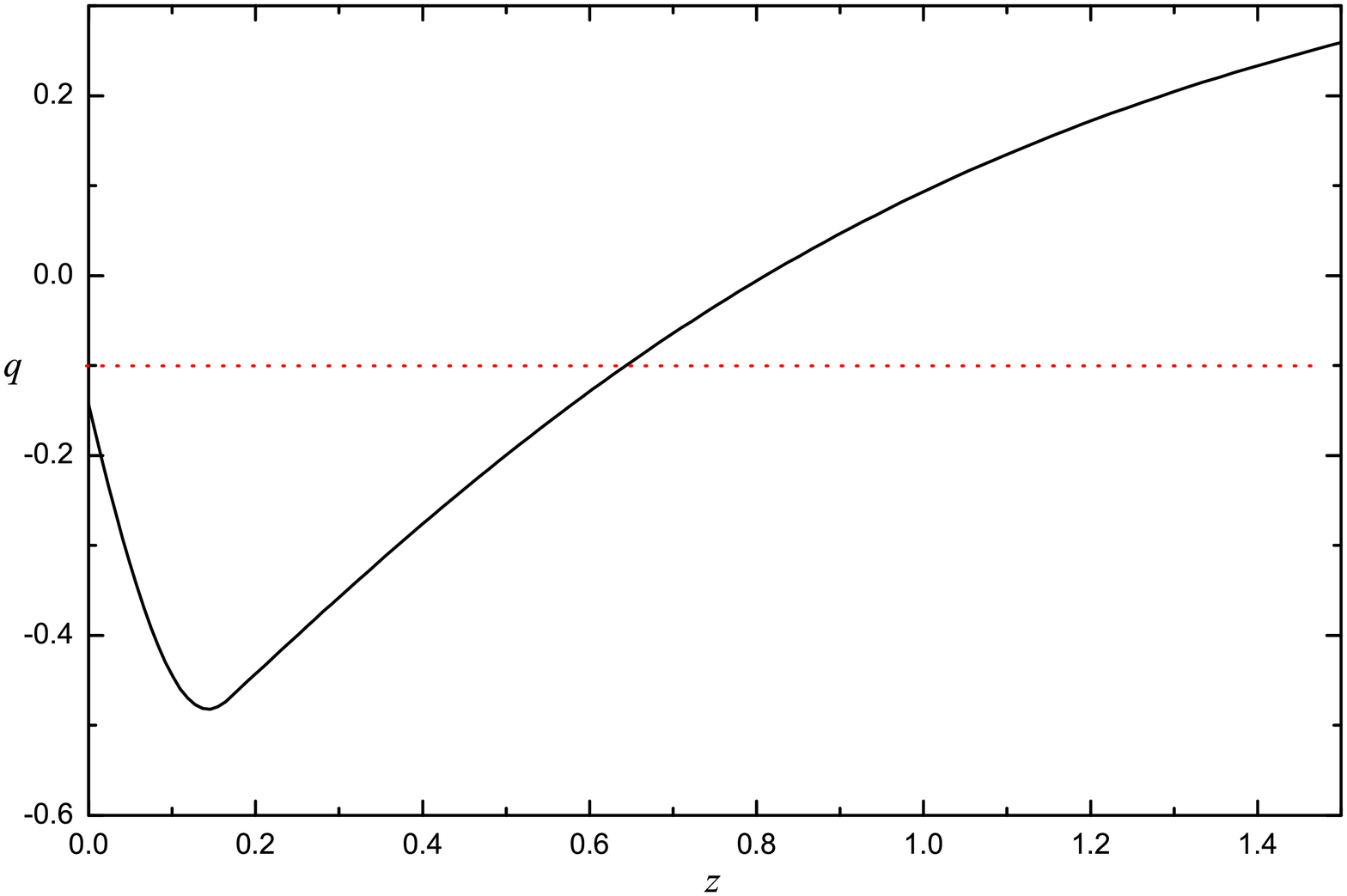}
\caption{Behavior of $\Omega_{\varphi}$ (dot line), 
$\Omega_{q}$ (dash  line) and $\Omega_{m}$ (solid line) as a function of  $N=\ln a$ for $n=3,\;\alpha = 0.005$ and $\mu = -5$ (upper side). Evolution of deceleration parameter for this model (lower side).}
\label{fig:4}
\end{figure}

In figure \ref{fig:4}, we show the evolution of $\Omega_q, \Omega_{m}$ and $\Omega_{\varphi}$ for  cosmological model  in the case when $\alpha=0.005$,
 $\mu=-5$ and $n = 3.$

From the form of equations and the character of the interaction it can be easily understood that neither the nature nor the location of the attractor, that have been found in the previous subsection dos not change with the inclusion of the interaction. Interaction only affects the behavior of dynamical variables  that are the correspond to trajectory  in phase space  between critical points. This is a consequence of the above degeneracy from the parameters of interaction.

With these values of the parameters of interaction, transient acceleration begins almost in the present era. In our model as in most cosmological models, where the scalar field plays the role of dark energy it begins to dominate causing a period of accelerated expansion of the Universe. Its consequence of the accelerated expansion of the Universe contribution of ADE increases, resulting that the background (space) is changing faster than the field and it becomes asymptotically free. This field has extremely rigid equation of state that leads to the fact that the accelerated expansion of the Universe is slowing down. Soon, however, when the contribution of ADE has grows enough so that the scalar field cannot longer impede the expansion of the Universe, it begins to accelerate again.

\section{Observational data}\label{OBS}
In the present section, we will consider the latest  cosmological observations, namely, the 557 SNIa sample (Union2) was released by the Supernova Cosmology Project (SCP) Collaboration~\cite{Amanullah}. 

The data points of the 557 Constitution SNIa compiled  in~\cite{Amanullah} are given in terms of the distance modulus  $\mu_{obs}(z_i)$. On the other hand, the theoretical  distance modulus is defined as
\begin{equation}
     \mu_{th}(z_i)\equiv 5\log_{10}D_L(z_i)+\mu_0\,,
\end{equation}
where $\mu_0\equiv 42.38-5\log_{10}h$ and $h$ is the Hubble  constant $H_0$ in units of $100~{\rm km\,s^{-1}\,Mpc^{-1}}$, whereas
\begin{equation}
     D_L(z)=(1+z)\int_0^z \frac{d\tilde{z}}{E(\tilde{z};{\bf p})}\,,
\end{equation}
in which $E\equiv H/H_0$, and ${\bf p}$ denotes the model parameters. 

\begin{figure}[t]
\centering
\includegraphics[width=0.45\textwidth]{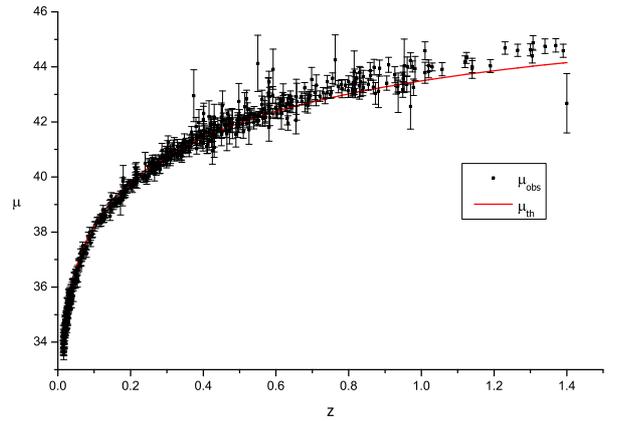}
\caption{The dependence of the modulus distance from the redshift, theoretically calculated (solid red line) by the model with  $Q=3H\alpha \rho_\varphi$, $\alpha=0.005$, $\mu=-5$ and $n=3$.  $ h = 0.70 $ $\rm \, km  \,c^{-1}\,Mpc^{-1} $  points found in the  observations of supernovae type 1a\cite{Amanullah}}.
\label{fig:mu_z_1}
\end{figure}

Theoretical distance modulus  will be different for the different models and comparing $\mu_{th}(z_i)$ with $\mu_{obs} (z_i) $, one can judge the plausibility of an cosmological model. So  to understand whether the theoretical model corresponds to the  observational data is enough to know the value of $E\equiv H/H_0$, which is easy to find through a system of equations (\ref{sys_xyz_V}). As seen from figure \ref{fig:mu_z_1} our  models are in accordance with the observational data. 

\section{Conclusion}
The original agegraphic dark energy model was proposed in~\cite{0707.4049} based  on the K\'{a}rolyh\'{a}zy uncertainty relation, which arises  from quantum mechanics together with general relativity.  The interacting agegraphic dark energy model has certain advantages compared to the original agegraphic or holographic dark energy model. Many studies show that this model gives an opportunity to explain the accelerated expansion of the Universe without a cosmological constant or some form of the scalar field. All the three models give dynamics of the Universe which are virtually indistinguishable from SCM, but without most of its problems, such as the cosmological constant, fine tuning and coincidence problems.

Some authors have recently suggested that the cosmic acceleration have already peaked and that we are currently observing its slowing down \cite{Barrow,Starobinsky,Lima}.
Under a kinematic analysis of the most recent SNe Ia compilations, the paper \cite{Lima,LiWuYu,LiWuYu1}  shows  the existence of a considerable probability in the relevant parameter space that the Universe is already in a decelerating expansion regime. 

One of the deficiencies of original ADE model is the inability to explain the phenomenon of transient acceleration. 

Density of holographic dark energy is determined by the surface terms in action, while volume terms are usually ignored. We take into account both surface and volume terms, where the latter correspond to (described by) homogeneous scalar field with exponential potential $V(\varphi)$.

We consider a model of Universe consisting of dark matter interacting with a scalar field on the agegraphic background. It is shown that this model can explain the transient acceleration. This model also is in accordance with the observational data.

\section*{Acknowledgements}

We are grateful to  Prof. Yu.L. Bolotin for kind help and discussions. We also thank V.A. Cherkaskiy for careful reading and editing of this article.

\bibliographystyle{model1-num-names}

\end{document}